\documentclass[10pt, final, journal, twocolumn]{IEEEtran}
\usepackage{amssymb}
\usepackage{mathtools}
\usepackage{graphicx}
\usepackage{cite}
\usepackage{soul}

\newcommand{\dB}{\mathrm{dB}\,}
\newcommand{\PU}{\mathrm{PU}\,}
\newcommand{\on}{\mathrm{on}\,}
\newcommand{\off}{\mathrm{off}\,}
\newcommand{\at}{\mathrm{at}\,}

\newcommand\blfootnote[1]{%
  \begingroup
  \renewcommand\thefootnote{}\footnote{#1}%
  \addtocounter{footnote}{-1}%
  \endgroup
}

\ifCLASSINFOpdf
\else
\fi
\begin{document}
%
\title{Extended Delivery Time Analysis for Cognitive Packet Transmission with Application to Secondary Queuing Analysis}
%
%
%

\author{Muneer Usman,~\IEEEmembership{Student Member,~IEEE,}
        Hong-Chuan Yang,~\IEEEmembership{Senior Member,~IEEE,}
        Mohamed-Slim Alouini,~\IEEEmembership{Fellow,~IEEE}
}

\maketitle

\begin{abstract}
Cognitive radio transceiver can opportunistically access the underutilized spectrum resource of primary systems for new wireless services. With interweave implementation, the secondary transmission may be interrupted by the primary user's transmission. To facilitate the delay analysis of such secondary transmission for a fixed-size secondary packet, we study the resulting extended delivery time that includes both transmission time and waiting time. In particular, we derive the exact distribution functions of extended delivery time of secondary transmission for both continuous sensing and periodic sensing cases. Selected numerical and simulation results are presented for illustrating the mathematical formulation. Finally, we consider an M/G/1 queueing set-up at the secondary transmitter and formulate the closed-form expressions for the expected delay with Poisson traffic. The analytical results will greatly facilitate the design of the secondary system for particular target application.
\end{abstract}


%
\IEEEpeerreviewmaketitle

\blfootnote{Manuscript received June 9, 2014; revised October 7, 2014 and February 16, 2015; accepted May 13, 2015. The associate editor coordinating the review of this paper and accepting it for publication is Giuseppe Bianchi.

Muneer Usman is associated with Google Inc, Mountain View, USA. (e-mail: muneer.usman@gmail.com)

Dr. Hong-Chuan Yang is associated with the Department of Electrical and Computer Engineering at the University of Victoria, BC, Canada (e-mail: hy@uvic.ca)

Dr. Mohamed-Slim Alouini is associated with the Computer, Electrical and Mathematical Science and Engineering Division at King Abdullah University of Science and Technology,  (e-mail: slim.alouini@kaust.edu.sa)

This work was supported in part by the Qatar National Research Fund (a member of Qatar Foundation) under NPRP Grant NPRP 5-250-2-087 and Discovery Grant from NSERC Canada. The statements made herein are solely the responsibility of the authors
}

\section{Introduction}
%
%
%
%

\IEEEPARstart{R}{adio} spectrum scarcity is one of the most serious problems nowadays faced by the wireless communications industry. Cognitive radio is a promising solution to this emerging problem by exploiting temporal/spatial spectrum opportunities over the existing licensed frequency bands \cite{Haykin,Mitola,Thomas,Akyildiz,Islam,Hamdaoui,Qianchuan,Qing}. Different implementation strategies exist for opportunistic spectrum access (OSA). In underlay cognitive radio implementation, the primary and secondary users simultaneously access the same spectrum, with a constraint on the interference that secondary user (SU) may cause to primary transmission. With interweave cognitive implementation, the secondary transmission creates no interference to the primary user (PU). Specifically, SU can access the spectrum only when it is not used by PU and must vacate the occupied spectrum when PU appears. Spectrum handoff procedures are adapted for returning the radio channel to the PU and then re-accessing that channel or another channel later to complete the transmission. As such, the secondary transmission of a given amount of data may involve multiple spectrum handoffs, which results in extra transmission delay. The total time required for the SU to complete a given packet transmission will include the waiting periods before accessing the channel and become more than the actual time needed for transmission. In this paper, we investigate the statistical characteristics of the resulting extended delivery time (EDT) \cite{Borgonovo} and apply them to evaluate the delay performance of secondary transmission.

\subsection{Previous Work}

There has been a continuing interest in the delay and throughput analysis for secondary systems, especially for underlay implementation. \cite{Khan} analyzes the delay performance of a point-to-multipoint secondary network, which concurrently shares the spectrum with a point-to-multipoint primary network in the underlay fashion, under Nakagami-$m$ fading. The packet transmission time for secondary packets under PU interference is investigated in \cite{Sibomana}, where multiple secondary users are simultaneously using the channel. An optimum power and rate allocation scheme to maximize
the effective capacity for spectrum sharing channels under average interference constraint is proposed in \cite{Musavian}. \cite{Tran} examines the probability density function (PDF) and cumulative distribution function (CDF) of secondary packet transmission time in underlay cognitive system. \cite{Farraj} investigates the M/G/1 queue performance of the secondary packets under the PU outage constraint. \cite{Jiang} analyzes the interference caused by multiple SUs in a ``mixed interweave/underlay'' implementation, where each SU starts its transmission only when the PU is off, and continues and completes its transmission even after the PU turns on.

For interweave implementation strategy, \cite{Gaaloul} discusses the average service time for the SU in a single transmission slot and the average waiting time, i.e. the time the SU has to wait for the channel to become available, assuming general primary traffic model. A probability distribution for the service time available to the SU during a fixed period of time was derived in \cite{Liang}. A model of priority virtual queue is proposed in \cite{J_Wang} to evaluate the delay performance for secondary users. \cite{Kandeepan} studies the probability of successful data transmission in a cooperative wireless communication scenario with hard delay constraints. A queuing analysis for secondary users dynamically accessing spectrum in cognitive radio systems was carried out in \cite{Li_Han}, where the authors analyze the maximum possible packet arrival rate the secondary system can support. \cite{Kahvand} proposes a dynamic channel selection approach to minimize the delay for secondary users in a pre-emptive resume priority M/G/1 queuing network. End-to-end performance of a cognitive radio network has been analyzed in terms of quality of service parameters in \cite{Namanya}.

The concept of EDT was first used in \cite{Borgonovo} to derive bounds on the throughput and delay performance of secondary users in cognitive scenario. The expected EDT of a packet for a cognitive radio network with multiple channels and users has been calculated in \cite{W_Wang}. \cite{W_Wang2} studies the EDT while taking spectrum sensing errors into account. In particular, the EDT statistics also depend on packet transmission strategy used when the secondary transmission is interrupted by PU activities, the secondary system can adopt either non-work-preserving strategy, where interrupted packets transmission must be repeated \cite{Borgonovo}, or work-preserving strategy, where the secondary transmission can continue from the point where it was interrupted, without wasting the previous transmission \cite{W_Wang, W_Wang2}. The work-preserving strategy can be achieved with the application of rateless codes such as Fountain code \cite{MacKay,Castura}. Work-preserving strategy also applies to the transmission scenario with small and individually coded sub-packets transmission.

\subsection{Contribution}

 In this paper, we carry out a thorough statistical analysis on the EDT of secondary packet transmission with work-preserving strategy. In general, the transmission of a secondary packet involves an interleaved sequence of transmission and waiting time slots, such that a transmission slot is followed by a waiting slot, followed by a transmission slot, and so on, each of which can have random time duration. We first derive the exact closed-form expression for the distribution function of EDT assuming a fixed packet transmission time, which might result from packet transmission over fast-varying fading channels, or static channels. Both the ideal scenario of continuous sensing, in which the SU will continuously sense for the channel availability, and the practical scenario of periodic sensing, in which the SU will sense the channel periodically, are considered. We also generalize the analysis to the case where the transmission time depends on the instantaneous channel quality, and as such, is random. The exact statistics of the EDT for secondary packet transmission can be directly used to predict the delay performance of some low-traffic intensity secondary applications.

We then apply these statistical results on EDT to the secondary queuing analysis. The queuing analysis for secondary transmission is a challenging problem even for Poisson arrival traffic. The main difficulty results from the fact that packets will experience two different types of service time depending on whether the packets see an empty queue upon arrival or not. In this paper, we solve this problem by using the mean value technique with the M/G/1 queuing model. The approach used is similar to M/G/1 queue analysis for machines with setup times \cite{Adan}.  Both average queuing delay and average queue length are calculated in closed form. Monte-Carlo simulations were carried out using Matlab to validate the analytical results. The queuing analysis can be used in design and performance analysis of secondary systems in a heavy traffic environment. 

Following is a summary of the major novel contributions of this paper:
\begin{enumerate}
	\item{Detailed delivery time formulation and analysis for secondary packet transmission with the consideration of periodic sensing. There has been little previous work on cognitive radio performance analysis with periodic sensing except for \cite{Gaaloul} which focuses on a single secondary transmission slot. \cite{Liu,Mariani} also consider periodic sensing but their focus is on design of periodic sensing parameters. We consider the scenario that packet transmission time, constant or random depending on fading channel conditions, spans over multiple secondary transmission slots under work-preserving strategy.}
	\item{Complete and exact statistics of the EDT for secondary packet transmission for both continuous sensing and periodic sensing scenarios. Most previous work on cognitive radio systems carry out the delay analysis by calculating the moments of the delivery time in various settings. To the best of our knowledge, the existing literature only studies the first two moments of EDT \cite{Borgonovo,W_Wang,W_Wang2}. We derive the exact distribution functions of EDT.}
	\item{ Accurate queuing analysis of secondary packet transmission considering two different service time characteristics for arriving packets depending on buffer status. We demonstrate that the conventional approach simply using the overall moments of service time will lead to inaccurate estimation of the actual average total delay. Other approaches have been used for secondary queuing analysis \cite{J_Wang,Li_Han}, but they are not readily applicable for periodic sensing scenario and/or with work-preserving strategy.}
\end{enumerate}

The rest of this paper is organized as follows. In section \ref{SystemModel}, we introduce the system model and the problem formulation. In section \ref{SecLT}, we analyze the EDT of a single packet for both continuous sensing case and periodic sensing case. We also consider the case of variable packet transmission time. In section \ref{SecHT}, we analyze the average queuing delay of the secondary system in a general M/G/1 queuing set-up. Finally, this paper is concluded in section \ref{Conclusion}.

\section{System Model and Problem Formulation}
\label{SystemModel}

We consider a cognitive transmission scenario where the SU opportunistically accesses a channel of the primary system for data transmission. The occupancy of that channel by the PU evolves independently according to a homogeneous continuous-time Markov chain with an average busy period of ${\lambda}$ and an average idle period of ${\mu}$\footnote{For mathematical tractability, we assume a simplified model here. Other traffic models such as Poisson packet arrival \cite{W_Wang2,B_Wang,Zhang} will be addressed in future work.}. Thus, the duration of busy and idle periods are exponentially distributed. The SU opportunistically accesses the channel in an interweave fashion. Specifically, the SU can use the channel only after PU stops transmission. As soon as the PU restarts transmission, the SU instantaneously stops its transmission, and thus no interference is caused to the PU.

\begin{figure} 
\includegraphics[width=3.4 in]{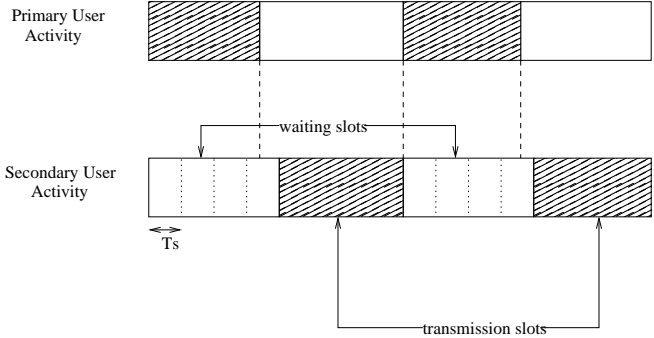}

\caption{Illustration of PU and SU activities and SU sensing for periodic sensing case.}
\label{cognitive_setup}
\end{figure}

The SU monitors PU activity through spectrum sensing\footnote{We assume perfect spectrum sensing here. The consideration of imperfect sensing will be treated in future work.}. Specific spectrum sensing methods have been discussed in detail in \cite{Akyildiz2,Cabric}, and here we focus on general sensing strategies. To ensure no interference to PU, we assume that during transmission, the SU always continuously monitors PU activity. However, when not transmitting, SU may adopt two sensing strategies i.e. continuous and periodic. With continuous sensing, the SU continuously senses the channel for availability. Thus, the SU starts its transmission as soon as the channel becomes available. As soon as the PU reappears, SU stops its transmission. We also consider the case where SU senses the channel periodically, with a period of $T_s$. If the PU is sensed busy, the SU will wait for $T_s$ time period and re-sense the channel. With periodic sensing, there is a small amount of time when the PU has stopped its transmission, but the SU has not yet acquired the channel, as illustrated in Fig. \ref{cognitive_setup}. During transmission, the SU continuously monitors PU activity. As soon as the PU restarts, the SU stops its transmission. The continuous period of time during which the PU is off and the SU is transmitting is referred to as a transmission slot. Similarly, the continuous period of time during which the PU is transmitting is referred to as a waiting slot. For periodic sensing case, the waiting slot also includes the time when the PU has stopped transmission, but the SU has not yet sensed the channel.

In this work, we analyze the packet delivery time of secondary system, which includes an interleaved sequence of the transmission slots and the waiting slots. The resulting EDT for a packet is mathematically given by $T_{ED}=T_w+T_{tr}$, where $T_w$ is the total waiting time for the SU and $T_{tr}$ is the packet transmission time. Note that both $T_w$ and $T_{tr}$ are, in general, random variables, with $T_w$ depending on $T_{tr}$, PU behaviour and sensing strategies, and $T_{tr}$, in turn, depending on packet size and secondary channel condition when available.   In what follows, we first derive the exact distribution of the EDT $T_{ED}$ for both continuous sensing and periodic sensing cases, which are then applied to the secondary queuing analysis in section \ref{SecHT}.

\section{Extended Delivery Time Analysis}
\label{SecLT}

In this section, we investigate the EDT of secondary system for a single packet arriving at a random point in time. We first consider the case where $T_{tr}$ can be viewed as a constant. This case applies to a fast varying channel, where the packet will experience different channel realizations over the duration of packet transmission. The transmission time $T_{tr}$ can be estimated as a constant depending on the ergodic channel capacity, given by \cite{Molisch}
\begin{equation}
T_{tr} \approx \frac{H}{W \int_0^{\infty} \log_2(1+{\gamma})f_{\gamma}(\gamma) d{\gamma} },
\end{equation}
where $H$ is the entropy of the packet, $W$ is the available bandwidth, and $f_{\gamma}(\gamma)$ is the PDF of the received SNR of the secondary channel. In case of a static channel, the transmission time will also be constant, given by
\begin{equation}
T_{tr} = \frac{H}{W \log_2(1+{\gamma_c})},
\end{equation}
where $\gamma_c$ is the constant SNR of the secondary channel. We then generalize the analysis to random $T_{tr}$ case by considering packet transmission over quasi-static channels for slow fading environment, where the received SNR  will remain constant during the transmission of a given packet and will be independent afterwards. The value of $T_{tr}$ will depend on the instantaneous SNR value during packet transmission. We also consider the case of very short packet transmission where the packet transmission will always complete in one secondary transmission slot. For both continuous sensing and periodic sensing scenarios, we derive the exact distribution of $T_{ED}$. These analyses also characterize the delay of some low-rate secondary applications. For example, in wireless sensor networks for health care monitoring, forest fire detection, air pollution monitoring, disaster prevention, landslide detection etc., the transmitter needs to periodically transmit measurement data to the sink with a long duty cycle. The EDT essentially characterizes the delay of measurement data collection.

\subsection{Continuous Sensing}
\label{SecLT_Cont}

\begin{figure} 
\includegraphics[width=3.4 in]{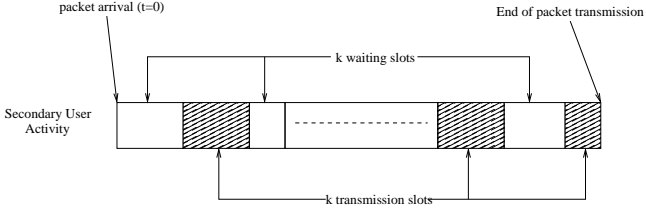}

\caption{Illustration of secondary transmission when the PU is on at the instant of packet arrival.}
\label{fn_k}
\end{figure}

\begin{figure} 
\includegraphics[width=3.4 in]{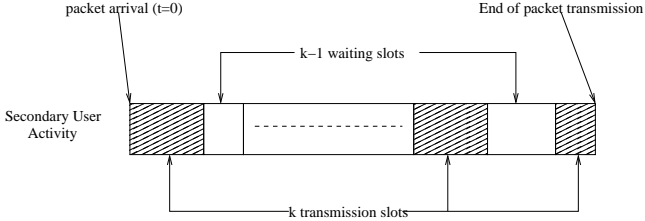}
\caption{Illustration of secondary transmission when the PU is off at the instant of packet arrival.}
\label{fn_k_1}
\end{figure}

The EDT for SU packet transmission consists of interleaved waiting slots and transmission slots. We first focus on the distribution of total waiting time $T_w$. The distribution of $T_w$ depends on whether the PU was on or off at the instant of packet arrival, as illustrated in Figs. \ref{fn_k} and \ref{fn_k_1}. We denote the PDF of the waiting time of the SU for the case when PU is on at the instant of packet arrival, and for the case when PU is off at the instant of packet arrival, by $f_{{T_w},p_{on}}(t)$ and $f_{{T_w},p_{off}}(t)$, respectively. The PDF of the waiting time $T_w$ for the SU is then given by
\begin{equation}
f_{T_w}(t) = \frac{\lambda}{{\lambda}+{\mu}} f_{{T_w},p_{on}}(t) + \frac{\mu}{{\lambda}+{\mu}} f_{{T_w},p_{off}}(t),
\label{ftw}
\end{equation}
where $\frac{\lambda}{{\lambda}+{\mu}}$ and $\frac{\mu}{{\lambda}+{\mu}}$ are the stationery probabilities that the PU is on or off at the instant of packet arrival, respectively. The two PDFs $f_{{T_w},p_{on}}(t)$ and $f_{{T_w},p_{off}}(t)$ above are calculated independently as follows.

When the PU is on at the instant of packet arrival, the number of transmission slots is same as the number of waiting slots. Hence, $T_w$ includes $k$ waiting slots if $k$ transmission slots are needed for packet transmission. Let ${{\cal{P}}_k}$ represent the probability that the SU completes packet transmission in exactly $k$ transmission slots, and $f_{{T_w},k}(t)$ represent the PDF of the total time duration of $k$ SU waiting slots. Note that due to the memoryless property of exponential distribution, ${{\cal{P}}_k}$ is independent of the distribution of the wait time $T_w$. Then the PDF of the total waiting time for the SU, for the case when PU is on at the instant of packet arrival, is given by
\begin{equation}
f_{{T_w},p_{on}}(t)=\sum_{k=1}^{\infty}{{{\cal{P}}_k} \times f_{{T_w},k}(t)}.
\label{ftw_pon}
\end{equation}
Note that for each value of $k$ in the summation, $T_w$ is the total time duration of $k$ waiting slots, each of which is exponentially distributed. Therefore $f_{{T_w},k}(t)$ is the PDF of the sum of $k$ independent and identically distributed exponential random variables with average $\lambda$, given by
\begin{equation}
f_{{T_w},k}(t) = \frac{1}{\lambda^{k}}\frac{t^{k-1}}{(k-1)!}e^{\frac{-t}{\lambda}}.
\label{ftw_k}
\end{equation}
To calculate ${{\cal{P}}_k}$, let us denote with ${{\cal{P}}^{(c)}_k}$ the probability that the SU completes its transmissions in $k$ or less transmission slots, which is the same as the probability that the total duration of $k$ transmission slots is more than $T_{tr}$. Since the total time for $k$ transmission slots, each of which follows exponential distribution, follows the Erlang distribution with PDF
\begin{equation}
f_{T_{tr},k}(t) = {\frac{1}{\mu^{k}}\frac{t^{k-1}}{(k-1)!}e^{\frac{-t}{\mu}}},
\end{equation}
${{\cal{P}}^{(c)}_k}$ can be calculated as the integral of Erlang PDF, given by
\begin{equation*}
{{\cal{P}}^{(c)}_k} = \int_{T_{tr}}^\infty{\frac{1}{\mu^{k}}\frac{t^{k-1}}{(k-1)!}e^{\frac{-t}{\mu}}} dt.
\end{equation*}
${{\cal{P}}_k}$ can then be calculated as
\begin{align}
\nonumber
{{\cal{P}}_k} &=  {{\cal{P}}^{(c)}_k} - {{\cal{P}}^{(c)}_{k-1}} \\
& = \int_{T_{tr}}^\infty{\frac{1}{\mu^{k}}\frac{t^{k-1}}{(k-1)!}e^{\frac{-t}{\mu}}} dt - \int_{T_{tr}}^\infty{\frac{1}{\mu^{k-1}}\frac{t^{k-2}}{(k-2)!}e^{\frac{-t}{\mu}}} dt.
\end{align}
After using integration by parts on the first integral and cancelling the terms, we obtain ${{\cal{P}}_k}$ as
\begin{equation}
{{\cal{P}}_k} = \frac{{T_{tr}}^{k-1}e^{\frac{-T_{tr}}{\mu}}}{\mu^{k-1}{(k-1)!}}.
\label{pr_nk}
\end{equation}
After substituting Eqs. (\ref{ftw_k}) and (\ref{pr_nk}) into Eq. (\ref{ftw_pon}), we get
\begin{equation}
f_{{T_w},p_{on}}(t)=\sum_{k=1}^{\infty}{  {\frac{{T_{tr}}^{k-1}e^{\frac{-{T_{tr}}}{\mu}}}{\mu^{k-1}{(k-1)!}}} \times {\frac{1}{\lambda^{k}}\frac{t^{k-1}}{(k-1)!}e^{\frac{-t}{\lambda}}} }.
\label{ftw_pon_summation}
\end{equation}
Finally, applying the definition of Bessel function, we arrive at the following closed-form expression for $f_{{T_w},p_{on}}(t)$
\begin{equation}
f_{{T_w},p_{on}}(t)=\frac{1}{\lambda}e^{\frac{-{T_{tr}}}{\mu}}{I_0 \left( 2{\sqrt{\frac{{T_{tr}}t}{\mu\lambda}}} \right)}{e^{\frac{-t}{\lambda}}},
\label{ftw_pon_final}
\end{equation}
where $I_n(.)$ is the modified Bessel function of the first kind of order $n$.

Similarly, noting that in case of PU off at the instant of packet arrival, the number of transmission slots is one more than the number of waiting slots, the PDF for $T_w$ when PU is off at the instant of packet arrival can be obtained as
\begin{align}
\nonumber
&f_{{T_w},p_{off}}(t) =\sum_{k=1}^{\infty}{{{\cal{P}}_k} \times f_{{T_w},k-1}(t)} \\ 
& = {e^{\frac{-{T_{tr}}}{\mu}}} {\delta}(t) + \sum_{k=2}^{\infty}{  {\frac{{T_{tr}}^{k-1}e^{\frac{-{T_{tr}}}{\mu}}}{\mu^{k-1}{(k-1)!}}} \times {\frac{1}{\lambda^{k-1}}\frac{t^{k-2}}{(k-2)!}e^{\frac{-t}{\lambda}}} },
\label{ftw_poff_summation}
\end{align}
which simplifies to 
\begin{equation}
f_{{T_w},p_{off}}(t) = {e^{\frac{-{T_{tr}}}{\mu}}} {\delta}(t) + \sqrt{ \frac{T_{tr}}{{\mu}{\lambda} t} } e^{\frac{-{T_{tr}}}{\mu}}{I_1 \left( 2{\sqrt{\frac{{T_{tr}}t}{\mu\lambda}}} \right)}{e^{\frac{-t}{\lambda}}},
\label{ftw_poff_final}
\end{equation}
where ${\delta}(t)$ is the delta function. Note that the term ${e^{\frac{-{T_{tr}}}{\mu}}} {\delta}(t)$ corresponds to the the case that the number of waiting slots is equal to $0$.

\begin{figure}[htb]
\includegraphics[width=3.4 in] {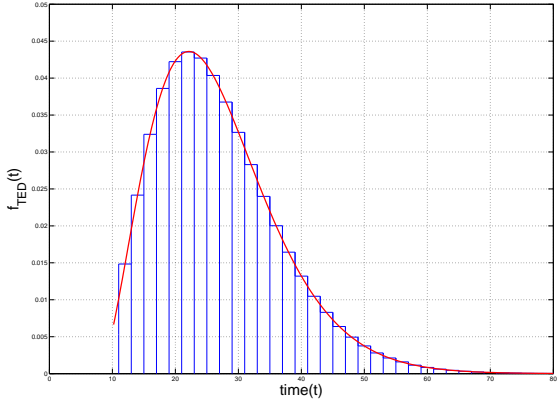}
\caption{Simulation verification for the analytical PDF of $T_{ED}$ with continuous sensing ($T_{tr} = 10$, $\lambda = 3$, and $\mu = 2$).}
\label{lt_cont_sim}
\end{figure}

After substituting Eqs. (\ref{ftw_pon_final}) and (\ref{ftw_poff_final}) into (\ref{ftw}), and noting $T_{ED} = T_w + T_{tr}$, the PDF for the EDT $T_{ED}$ for continuous sensing case is given by
\begin{align}
\nonumber
f_{T_{ED}}(t) & = u(t-{T_{tr}}) \frac{1}{{\lambda}+{\mu}}  {e^{\frac{-(t-T_{tr})}{\lambda}}} {e^{\frac{-T_{tr}}{\mu}}} \\
& \times \left[ {I_0 \left( 2{\sqrt{\frac{{T_{tr}}(t-{T_{tr}})}{\mu\lambda}}} \right)} \right. \\
\nonumber
&  \left. + \sqrt{ \frac{{T_{tr}} {\mu}}{{\lambda} (t-{T_{tr}})}} {I_1 \left( 2{\sqrt{\frac{{T_{tr}}(t-{T_{tr}})}{\mu\lambda}}} \right)} \right] \\
& + \frac{\mu}{{\lambda}+{\mu}} {e^{\frac{-{T_{tr}}}{\mu}}} {\delta}(t-{T_{tr}}),
\label{f_TED}
\end{align}
where $u(.)$ is the step function.

Fig. \ref{lt_cont_sim} plots the analytical expression for the PDF of the EDT with continuous sensing, given in Eq. (\ref{f_TED}). For the purpose of Monte-Carlo simulations, a cognitive radio system was simulated with randomly generated PU on and off times. A large number of packets was simulated, and the results were compiled to estimate the PDF of the EDT. The plot for the simulation results is also shown in Fig. \ref{lt_cont_sim}. The perfect match between analytical and simulation results verify our analytical approach.

\subsection{Periodic Sensing}
\label{SecLT_Per}
In the case of periodic sensing, the waiting time $T_w$ will be a multiple of $T_s$, which is a known constant quantity. Therefore $T_w$ will have a discrete distribution. Similar to continuous sensing case, we can write the probability that the waiting time $T_w$ is $nT_s$ by considering the PU is on or off at the instant of packet arrival separately, as
\begin{align}
\nonumber
\Pr[T_w = n T_s] = & \frac{\lambda}{{\lambda}+{\mu}} \Pr[T_w,p_{on} = n T_s] \\
& + \frac{\mu}{{\lambda}+{\mu}} \Pr[T_w,p_{off} = n T_s],
\label{pr_tw}
\end{align}
where $\Pr[T_w,p_{on} = n T_s]$ is the probability that the total waiting time for the SU is $n T_s$ when PU is on at the instant of packet arrival, and $\Pr[T_w,p_{off} = n T_s]$ the probability when PU is off at the instant of packet arrival.

It can be shown, based on the illustration in Fig. \ref{fn_k}, that
\begin{equation}
\Pr[T_w,p_{on} = n T_s] = \sum_{k=1}^{\infty}{ {{\cal{P}}_k} \times \Pr[T_{w,k} = n T_s]},
\label{pr_tw_pon}
\end{equation}
where ${{\cal{P}}_k}$ is the probability that the SU completes its transmission in $k$ slots, which is independent of sensing strategy and also given in Eq. (\ref{pr_nk}), and $\Pr[T_{w,k} = n T_s]$ is the probability that the SU waiting time in $k$ slots is $n T_s$, which is, as shown in Appendix \ref{app_beta}, given by\footnote{Here we assume perfect sensing. For the case of imperfect periodic sensing, this PMF will change, while most of the remaining formulation will remain the same.}
\begin{equation}
\Pr[T_{w,k} = n T_s] = {(1-\beta)^k}{(\beta)^{n-k}}{{n-1}\choose{k-1}},
\label{pr_twk}
\end{equation}
where $\beta$ is the probability that primary user is on at the sensing instant, which is also obtained in Appendix \ref{app_beta} as
\begin{equation} 
\beta=\frac{\lambda}{\lambda+\mu}+{\frac{\mu}{\lambda+\mu}}{e^{-(\frac{1}{\lambda}+\frac{1}{\mu}){T_s}}}.
\label{beta}
\end{equation}
Note that when $T_s$ is very small, Eq. (\ref{pr_twk}) converges to Eq. (\ref{ftw_k}), as shown in appendix \ref{app_cont_to_per_conv}. After substituting Eqs. (\ref{pr_nk}) and (\ref{pr_twk}) into Eq. (\ref{pr_tw_pon}), and some manipulation, we can calculate $\Pr[T_w,p_{on} = n T_s]$ as
\begin{align}
\nonumber
\Pr[T_w,p_{on} = n T_s] & = {(1-\beta)}{\beta^{n-1}}{e^{\frac{-{T_{tr}}}{\mu}}} \\
& \times {\sum_{0}^{n-1}{\left[\frac{{T_{tr}}(1-\beta)}{\mu\beta}\right]^k}{\frac{1}{k!}}{{n-1}\choose{k}}},
\label{pr_tw_pon_summation}
\end{align}
which simplifies to
\begin{align}
\nonumber
\Pr[T_w,p_{on} = n T_s] & = {(1-\beta)}{\beta^{n-1}}{e^{\frac{-{T_{tr}}}{\mu}}} \\
& \times {{}_1 F_1 \left(1-n;1;{\frac{-{T_{tr}}(1-\beta)}{\mu\beta}} \right)},
\label{pr_tw_pon_final}
\end{align}
where ${}_1 F_1(.,.,.)$ is the generalized Hyper-geometric function.
Similarly, $\Pr[T_w,p_{off} = n T_s]$ in Eq. (\ref{pr_tw}) can be calculated as
\begin{equation}
\Pr[T_w,p_{off} = n T_s] = \sum_{k=1}^{\infty}{{{\cal{P}}_k} \times \Pr[T_{w,k-1} = n T_s] }.
\label{pr_tw_poff}
\end{equation}
After substituting Eqs. (\ref{pr_nk}) and (\ref{pr_twk}) into Eq. (\ref{pr_tw_poff}), and some manipulation, we get
\begin{align}
\nonumber
& \Pr[T_w,p_{off} = n T_s] = {\left[\frac{{T_{tr}}(1-\beta){\beta}^{n-1}}{\mu}\right]} {e^{\frac{-{T_{tr}}}{\mu}}} \\
\nonumber
& \times \sum_{k=0}^{n-1}  {\left[\frac{{T_{tr}}(1-\beta)}{\mu\beta}\right]^k}  {\frac{1}{(k+1)!}}{{n-1}\choose{k}} \\
& + e^{-\frac{T_{tr}}{\mu}} {\delta}[n],
\end{align}
which eventually simplifies to
\begin{align}
\nonumber
& \Pr[T_w,p_{off} = n T_s] = {\left[\frac{{T_{tr}}(1-\beta){\beta}^{n-1}}{\mu}\right]} {e^{\frac{-{T_{tr}}}{\mu}}} \\
& \times {{}_1 F_1 \left(1-n;2;{\frac{-{T_{tr}}(1-\beta)}{\mu\beta}} \right) } u[n-1] + e^{-\frac{T_{tr}}{\mu}} {\delta}[n].
\label{pr_tw_poff_final}
\end{align}

After substituting Eqs. (\ref{pr_tw_pon_final}) and (\ref{pr_tw_poff_final}) into Eq. (\ref{pr_tw}), the probability mass function (PMF) of the EDT for periodic sensing case is given by
\begin{align}
\nonumber
& \Pr[ T_{ED} = n T_s + T_{tr}] = \frac{\lambda}{{\lambda}+{\mu}}  {(1-\beta)}{\beta^{n-1}}{e^{\frac{-{T_{tr}}}{\mu}}} \\
\nonumber
& \times{{}_1 F_1 \left(1-n;1;{\frac{-{T_{tr}}(1-\beta)}{\mu\beta}} \right)} u[n] \\
\nonumber
& + \frac{\mu}{{\lambda}+{\mu}} \left[ { \left(\frac{{T_{tr}}(1-\beta){\beta}^{n-1}}{\mu} \right)} {e^{\frac{-{T_{tr}}}{\mu}}} \right. \\
& \left. {{}_1 F_1 \left(1-n;2;{\frac{-{T_{tr}}(1-\beta)}{\mu\beta}} \right)} u[n-1] + e^{-\frac{T_{tr}}{\mu}} {\delta}[n] \right].
\label{pr_TED}
\end{align}

\begin{figure} 
\includegraphics[width=3.4 in]{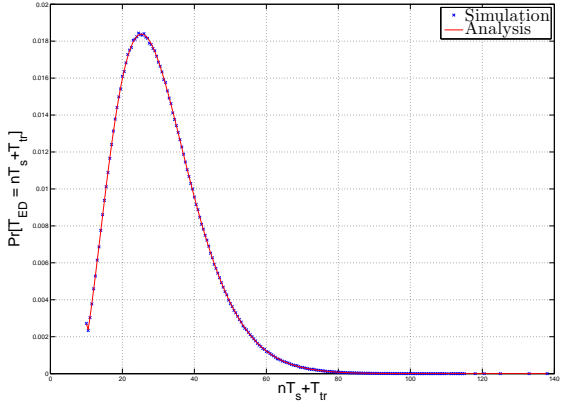}

\caption{Simulation verification of the analytical PMF of $T_{ED}$ with periodic sensing ($T_{tr} = 10$, $\lambda = 3$, $\mu = 2$, and $T_s=0.5$).}
\label{lt_per_sim}
\end{figure}

\begin{figure} 
\includegraphics[width=3.4in]{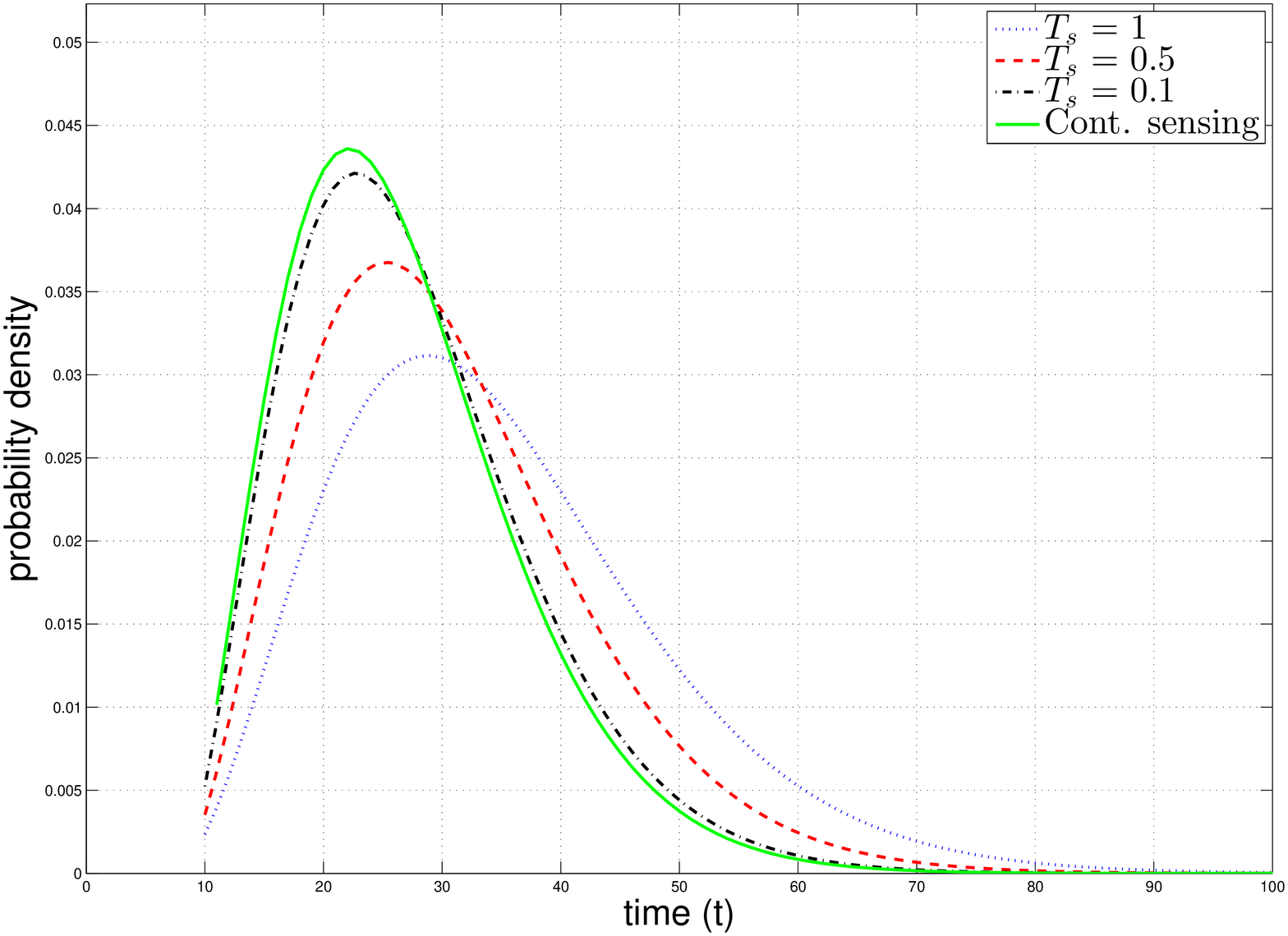}

\caption{Distribution of the EDT with continuous and periodic sensing ($T_{tr} = 10$, $\lambda = 3$, and $\mu = 2$).}
\label{lt_delay_dist}
\end{figure}

Fig. \ref{lt_per_sim} plots the PMF of extended delivery time $T_{ED}$ for periodic sensing case, and the corresponding simulation result. The plots show that the analytical results conform to the simulation results. Fig. \ref{lt_delay_dist} shows the PMF envelope of the packet delivery time with periodic sensing for various values of sensing period $T_s$. As can be seen, the performance of periodic sensing improves with reduction in the sensing interval $T_s$. As $T_s$ approaches $0$, the performance of periodic sensing converges to that of continuous sensing, as expected.

\subsection{Consideration of random transmission time}
\label{ShortPackets}

In the earlier subsections, the packet transmission time $T_{tr}$ was considered to be a constant, which applies to packet transmission over fast fading channel or static channel scenarios. Now we consider the transmission of packets over a slow fading channel, where $T_{tr}$ depends on the instantaneous SNR of the secondary channel.

\subsubsection{Constant SNR during packet transmission}

For packet transmission over slow and quasi-static fading channels, we assume that the received SNR of the secondary channel $\gamma$, which is a random variable, will not change for the complete duration of the packet transmission. Different packets will experience different channel conditions. The transmission time $T_{tr}$ now becomes a random variable, as it is a function of the received SNR $\gamma$, defined as
\begin{equation}
T_{tr} = \frac{H}{W \log_2(1+\gamma)},
\end{equation}
where $H$ is the entropy of the data packet in bits. Assuming a Rayleigh fading model for the secondary channel with an SNR PDF given by
\begin{equation}
f_{\gamma}(\gamma) = \frac{1}{\bar{\gamma}} e^{\frac{\gamma}{\bar{\gamma}}},
\end{equation}
the PDF of the transmission time $T_{tr}$ can be derived as \cite{Molisch}
\begin{equation}
f_{T_{tr}}(T) = \frac{H}{\bar{\gamma}T^2} e^{\left[ \frac{1}{\bar{\gamma}} + \frac{H}{T} - \frac{e^{\frac{H}{T}}}{\bar{\gamma}} \right]},
\label{pdf_Ttr}
\end{equation}
where $\bar{\gamma}$ is the average link SNR. The exact distribution of the EDT of a single packet for the continuous sensing case can then be calculated as
\begin{equation}
f_{T_{ED}}^{v,c}(t) = \int_0^{t} f_{T_{ED}}(t|T_{tr}) \cdot f_{T_{tr}}(T_{tr}) d{T_{tr}},
\end{equation}
where $f_{T_{ED}}(.|T_{tr})$ is the conditional PDF of the EDT of the SU for a given $T_{tr}$, as given in Eq. (\ref{f_TED}). For the periodic sensing case, the exact distribution of EDT will be given by
\begin{align}
\nonumber
f_{T_{ED}}^{v,p}(t) = \sum_{n=0}^{\lfloor \frac{t}{T_s} \rfloor} & \Pr[T_{ED} = n T_s + T_{tr} | T_{tr} = t-n T_s] \\
& \times f_{T_{tr}}(t-n T_s),
\end{align}
where $\Pr[T_{ED} = n T_s + T_{tr} | T_{tr} = t-n T_s]$ is the probability that the EDT of a packet for given data transmission time $T_{tr}$ is $n T_s + T_{tr}$, as defined in Eq. (\ref{pr_TED}). More specifically, each summation term in the above equation refers to the probability that the SU waiting time is $n T_s$ and the physical packet transmission time is $t - n T_s$.

\begin{figure} 
\includegraphics[width=3.4in]{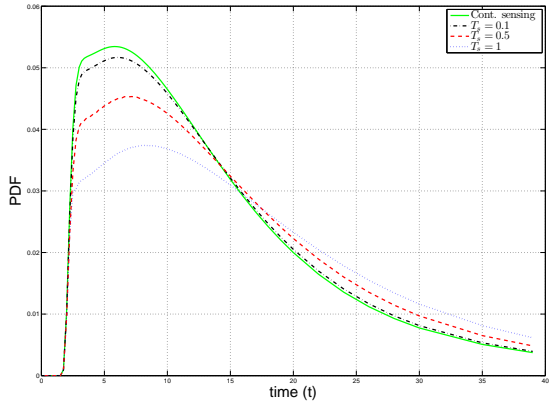}

\caption{PDF of EDT for short packets ($H = 100$, $W = 10$, $\bar{\gamma} = 8 \; \dB$ $\lambda = 3$, and $\mu = 2$).}
\label{special_case1}
\end{figure}

Fig. \ref{special_case1} shows the numerically computed PDFs of the EDT over slow fading channels for continuous sensing case and periodic sensing cases with $T_s = 0.1$, $T_s=0.5$, and $T_s = 1$. As the periodic sensing interval approaches $0$, the corresponding PDF curve approaches the PDF curve for continuous sensing.

\subsubsection{One shot transmission}

In certain sensing applications, the secondary packets can be very short. They may only contain a temperature measurement, or a meter reading. The transmission of such packets may complete in a single secondary transmission slot. In this subsection, we analyze the EDT of such packets. Note that in this case, the secondary packet needs to wait for at most one transmission slot. Using the PDF given Eq. (\ref{ftw_k}) with $k=1$, the PDF of the EDT for such packets with continuous sensing can be calculated as
\begin{align}
\nonumber
f_{T_{ED}}^{o,c}(t) & = \frac{\lambda}{{\lambda}+{\mu}} \int_{0}^{t} \frac{1}{\lambda}e^{\frac{-(t-T_{tr})}{\lambda}} f_{T_{tr}}(T_{tr}) dT_{tr} \\
& + \frac{\mu}{{\lambda} + {\mu}} f_{T_{tr}}(t).
\end{align}
Similarly, for periodic sensing case, using $\Pr[T_{w,k} = n T_s]$ from Eq. (\ref{pr_twk}) with $k=1$, the probability distribution function of the EDT for such packets is given by
\begin{align}
\nonumber
& f_{T_{ED}}^{o,p}(t) = \frac{\mu}{{\lambda}+{\mu}} f_{T_{tr}}(t) \\
& + \frac{\lambda}{{\lambda}+{\mu}} \sum_{n=1}^{\lfloor \frac{t}{T_s} \rfloor} {(1-\beta)}{(\beta)^{n-1}} \cdot f_{T_{tr}}(t-n.T_s). 
\label{f_od}
\end{align}

\begin{figure} 
\includegraphics[width=3.4in]{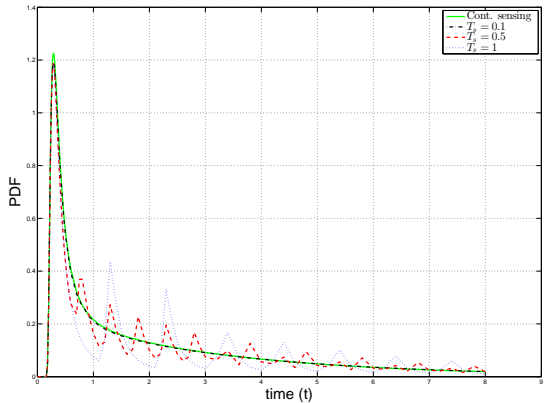}

\caption{PDF of EDT for one shot transmission ($H = 10$, $W = 10$, $\bar{\gamma} = 8 \; \dB$ $\lambda = 3$, and $\mu = 2$).}
\label{special_case2}
\end{figure}

Fig. \ref{special_case2} displays the  numerically computed PDFs of the EDT for very short packets for continuous sensing and periodic sensing cases. The first peak in all the curves correspond to the case that the incoming packet finds the PU to be off, and hence gets transmitted immediately. The oscillations seen in the curves for $T_s=0.5$ and $T_s=1$ can be attributed to sharp-peaked nature of the PDF of the transmission time $T_{tr}$, where each peak in the above curve corresponds to a different value of $n$ in Eq. (\ref{f_od}).

\section{Application to Secondary Queuing Analysis}
\label{SecHT}

\begin{figure} 
\includegraphics[width=3.4 in]{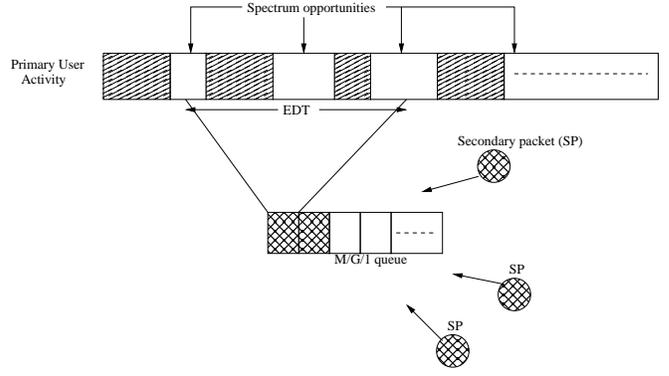}

\caption{M/G/1 queuing system model}
\label{queuing_model}
\end{figure}

In this section, we consider the transmission delay for the secondary system in a queuing set-up as illustrated in Fig. \ref{queuing_model}. In particular, the secondary traffic intensity is high and, as such, a first-in-first-out queue is introduced to hold packets until being transmitted. We assume that the arrival of equal-sized packet follows a Poisson process with intensity $\frac{1}{\psi}$, i.e. the average time duration between packet arrivals is ${\psi}$. For the sake of presentation clarity, the packets are assumed to be of the same length, with fixed constant transmission time $T_{tr}$ in the following analysis. The assumption of same long length is quite applicable to wireless networks, where data streams are typically fragmented into equal-size packets. As such, the secondary packet transmission can be modelled as a general M/G/1 queue, where the service time is closely related to the EDT that we studied in the previous section. The service time of a packet is defined as the time duration from the instant when the packet becomes available for transmission, either by arriving to an empty queue or by becoming the first packet in the queue due to the completion of previous packets' transmission, until the instant when it gets completely transmitted. Therefore, the service time also consists of interleaved sequence of transmitting and waiting slots.

Note from the EDT analysis, that the total waiting time of a packet depends on whether the PU is on or off when the packet is available for transmission. As such, different secondary packets will experience two types of service time characteristics. Specifically, some packets might see that there are one or more packets waiting in the queue or being transmitted upon arrival. Such packets will have to wait in the queue until transmission completion of previous packets. Once all the previous packets are transmitted, the new arriving packet will find the PU to be off. We term such packets as Type 1 packets. On the other hand, some packets will arrive when the queue is empty, and will immediately become available for transmission. Such packets might find the PU to be on or off. We will call this type of packets, Type 2 packets. To facilitate subsequent queuing analysis, we now calculate the first and second moments of the service time for these two types of packets.

\subsection{Moments of packet service time}
\label{HT_moments}

\subsubsection{Type 1 packets}
The average service time of Type 1 packets is equal to the EDT of packets that find PU off at the start of their transmission. Specifically, the first moment of the service time of Type 1 packets with continuous sensing can be calculated as
\begin{equation}
E[ST_{Type1}] = {\int}_{T_{tr}}^{\infty} t f_{{T_w},p_{off}}(t-{T_{tr}}) dt \triangleq E[ST_{p_{off}}]^{(c)}, 
\label{E_1_t_cont_def}
\end{equation}
where $E[ST_{p_{off}}]^{(c)}$ denotes the average EDT of a packet that finds PU off at the start of their transmission for continuous sensing. Substituting Eq. (\ref{ftw_pon_summation}) into Eq. (\ref{E_1_t_cont_def}) and carrying out integration while following the steps in Appendix \ref{app_HT_Cont}, we can obtain the closed-form expression of $E[ST_{Type1}]$ as
\begin{equation}
E[ST_{Type1}] = E[ST_{p_{off}}]^{(c)} = T_{tr} + {\lambda} \left({\frac{T_{tr}}{\mu}} \right).
\label{E1_t_cont}
\end{equation}
Similarly, the second moment of service time for Type 1 packet with continuous sensing can be calculated as
\begin{align}
\nonumber
E[ST_{Type1}^2] &= E[ST_{p_{off}}^2]^{(c)} 
\\ &= {{\lambda}^2} \left[ {\left(\frac{T_{tr}}{\mu} \right)}^2 + 2 {\frac{T_{tr}}{\mu}} \right] + 2{\frac{{\lambda}{T_{tr}}^2}{\mu}} + {T_{tr}}^2.
\end{align}

With periodic sensing, the first and second moment of service time of Type 1 packets can be calculated using $\Pr[T_w,p_{off} = n T_s]$ as
\begin{align}
\nonumber
E[ST_{Type1}] &= T_s \sum_{n=1}^{\infty} n Pr[T_w,p_{off} = n T_s] + T_{tr} 
\\ & \triangleq E[ST_{p_{off}}]^{(p)},
\end{align}
and
\begin{align}
\nonumber
& E[ST_{Type1}^2] = {T_s}^2 \sum_{n=1}^{\infty} n^2 Pr[T_w,p_{off} = n T_s]  \\
& + 2 T_s T_{tr} \sum_{n=1}^{\infty} n Pr[T_w,p_{off} = n T_s] + {T_{tr}}^2 \triangleq E[ST_{p_{off}}^2]^{(p)},
\end{align}
respectively, where $E[ST_{p_{off}}]^{(p)}$ and $E[ST_{p_{off}}^2]^{(p)}$ denote the average EDT of a packet that find PU off at the start of their transmission for periodic sensing. Following similar steps in Appendix \ref{app_HT_Per}, we can obtain the following closed-form expressions of $E[ST_{Type1}]$ and $E[ST_{Type1}^2]$ for periodic sensing case
\begin{equation}
E[ST_{Type1}] = E[ST_{p_{off}}]^{(p)} = {T_{tr}} \left(1+\frac{T_s}{\mu({1-{\beta}})} \right)
\label{E1_t_per}
\end{equation}
and
\begin{align}
\nonumber
E[ST_{Type1}^2] &= E[ST_{p_{off}}^2]^{(p)} = \frac{{T_s}^2}{{(1-{\beta})}^2} \left[ \left(\frac{T_{tr}}{\mu} \right)^2 + 2 {\frac{T_{tr}}{\mu}} \right] \\
&- \frac{{T_s}^2}{1-{\beta}} \left(\frac{T_{tr}}{\mu} \right) + \frac{{T_s}}{1-{\beta}} \frac{2 {T_{tr}}^2}{\mu}  + {T_{tr}}^2.
\end{align}

\subsubsection{Type 2 packets}
Type 2 packets may find PU on or off at the start of their service upon arrival. Therefore, the service time of Type 2 packets is the weighted average of the EDTs of packets that find PU on at the start of their transmission, and those that find PU off. Mathematically speaking, $E[ST_{Type2}]$ and $E[ST_{Type2}^2]$ can be calculated as
\begin{equation}
E[ST_{Type2}] = P_{on,2} \cdot E[ST_{p_{on}}] + (1-P_{on,2}) \cdot E[ST_{p_{off}}],
\label{E2_t}
\end{equation}
and
\begin{equation}
E[ST_{Type2}^2] = P_{on,2} \cdot E[ST_{p_{on}}^2] + (1-P_{on,2}) \cdot E[ST_{p_{off}}^2],
\label{E2_t2}
\end{equation}
where $P_{on,2}$ denotes the probability that a Type 2 packet finds PU on upon arrival, $E[ST_{p_{on}}]$ and $E[ST_{p_{on}}^2]$ are the first and second moments of the EDT of a packet that finds PU on at the instant of packet arrival, respectively, and $E[ST_{p_{off}}]$ and $E[ST_{p_{off}}^2]$ are the moments for PU off case. In particular, $E[ST_{p_{on}}]$ and $E[ST_{p_{on}}^2]$ have been calculated for continuous sensing case in Appendix \ref{app_HT_Cont}, and for periodic sensing case in Appendix \ref{app_HT_Per}.

We now derive an expression for $P_{on,2}$ through the following argument. Whenever the transmission of the last packet in the queue is completed, due to the memoryless property of exponential distribution, the time that takes for the next packet to arrive will follow an exponential distribution with average $\psi$. At the start of that time interval, it is known that the PU is off. The probability that the PU is on, $P_{p_{on}}(t)$, conditioned on the time elapsed since the completion of last packet transmission, $t$, is given by \cite{Cinlar}
\begin{align}
\nonumber
P_{p_{on}}(t) &= \Pr[\PU \; \on \; \at \; t_0+t \; | \; \PU \; \off \; \at \; t_0] \\
&= \frac{\lambda}{\lambda + \mu} \left[ 1 - e^{-(\frac{1}{\lambda} + \frac{1}{\mu})t} \right].
\end{align}
Removing the conditioning on $t$, the probability for the PU being on when a Type 2 packet arrives, $P_{on,2}$, is obtained as
\begin{equation}
P_{on,2} = E[P_{p_{on}}(t)] = \int_0^{\infty} P_{p_{on}}(t) \cdot \frac{1}{\psi} e^{-\frac{t}{\psi}} dt.
\end{equation}
It can be shown that the above simplifies to
\begin{equation}
P_{on,2} = \frac{\lambda \psi}{\lambda \psi + \lambda \mu + \mu \psi}.
\label{P_on_2}
\end{equation}

Substituting the moments $E[ST_{p_{on}}]$, $E[ST_{p_{on}}^2]$, $E[ST_{p_{off}}]$, and $E[ST_{p_{off}}^2]$ for continuous and periodic sensing cases, and $P_{on,2}$ into Eqs. (\ref{E2_t}) and (\ref{E2_t2}), we can obtain the moments of Type 2 packet service time. As an example, the first moment of the service time for Type 2 packets with continuous sensing is given, after substituting Eqs. (\ref{E1_t_cont}), (\ref{P_on_2}), and (\ref{et_on_cont}) into Eq. (\ref{E2_t}), by
\begin{equation}
E[ST_{Type2}] = \frac{\lambda^2 \psi}{\lambda \psi + \lambda \mu + \mu \psi} + \left(1+\frac{\lambda}{\mu} \right)T_{tr}.
\end{equation}
The other moments can be similarly obtained, but omitted here for conciseness.

\subsection{Queuing Analysis}

In this subsection, we derive an expression for the expected delay for a packet in the queue. For clarity, we focus on continuous sensing in the following. The expression for periodic sensing can be similarly obtained. The average total delay is given by
\begin{equation}
E[D] = E[ST] + E[Q],
\end{equation}
where $E[ST]$ is the average service time of an arbitrary packet, and $E[Q]$ is the average wait time in the queue. $E[ST]$ is a weighted average of $E[ST_{Type1}]$ and $E[ST_{Type2}]$, as defined in Eqs. (\ref{E1_t_cont}) and (\ref{E2_t}), respectively, given by
\begin{equation}
E[ST] =  (1-p_0) \cdot E[ST_{Type1}] + p_0 \cdot E[ST_{Type2}] ,
\label{simultaneous_eq_1}
\end{equation}
where $p_0$ is the probability of the queue being empty at any given time instance and $1-p_0$ is the utilization factor of the queue, which is, in turn, related to $E[ST]$ as
\begin{equation}
1-p_0 = \frac{E[ST]}{\psi}.
\label{simultaneous_eq_2}
\end{equation}
Simultaneously solving Eqs. (\ref{simultaneous_eq_1}) and (\ref{simultaneous_eq_2}), we can obtain $E[ST]$ and $p_0$ as
\begin{equation}
E[ST] = \frac{\psi E[ST_{Type2}]}{\psi + E[ST_{Type2}] - E[ST_{Type1}]},
\end{equation} 
and
\begin{equation}
p_0 = \frac{\psi - E[ST_{Type1}]}{\psi + E[ST_{Type2}] - E[ST_{Type1}]},
\label{p0}
\end{equation} 
respectively.

The average delay in the queue, $E[D]$, can be calculated using the mean value technique \cite{Boucherie} as
\begin{equation}
E[Q] = E[N_Q] \cdot E[ST_{Type1}] + (1-p_0) \cdot E[R],
\label{E_W}
\end{equation}
where $E[N_Q]$ is the average number of packets waiting in the queue, not including the current packet in service, $E[ST_{Type1}]$ is the average service time of a packet in the queue (Type 1 packet), and $E[R]$ is the mean residual time of the packet currently being served. Specifically, the first addition term in Eq. (\ref{E_W}) corresponds to the average total service time of the packets currently waiting in the queue, if any, and the second term to the waiting time for the currently served packet, if any. Given that a packet is being served at a given instance, the probabilities that the packet is a Type 1 packet or Type 2 packet, are equal to $\frac{(1-p_0) E[ST_{Type1}]}{(1-p_0) E[ST_{Type1}] + p_0 E[ST_{Type2}]}$ and $\frac{p_0 E[ST_{Type2}]}{(1-p_0) E[ST_{Type1}] + p_0 E[ST_{Type2}]}$, respectively. Therefore, mean residual service time, $E[R]$ can be calculated as
\begin{align}
\nonumber
E[R] & = \frac{(1-p_0) E[ST_{Type1}] }{(1-p_0) E[ST_{Type1}] + p_0 E[ST_{Type2}]} \cdot E[R_1] \\
& + \frac{p_0 E[ST_{Type2}]}{(1-p_0) E[ST_{Type1}] + p_0 E[ST_{Type2}] } \cdot E[R_2], 
\label{E_R}
\end{align}
where $E[R_1]$ and $E[R_2]$ are the mean residual times for Type 1 and Type 2 packets, respectively, defined by \cite{Adan}
\begin{equation}
E[R_1] = \frac{E[ST_{Type1}^2]}{2 E[ST_{Type1}]},
\end{equation}
and
\begin{equation}
E[R_2] = \frac{E[ST_{Type2}^2]}{2 E[ST_{Type2}]}.
\end{equation}
Recalling the Little's law stating that
\begin{equation}
E[N_Q] = \frac{E[Q]}{\psi},
\label{E_N_Q}
\end{equation} 
$E[Q]$ can be obtained after substituting Eqs. (\ref{p0}), (\ref{E_R}), and (\ref{E_N_Q}) into Eq. (\ref{E_W}) and much simplification as
\begin{equation}
E[Q] = \frac{ E[ST^2]}{2({\psi}-E[ST_{Type1}])},
\end{equation}
where $E[ST^2]$ is the second moment of the average service time of all packets, and is calculated as weighted sum of the second moments of the two packet types as
\begin{align}
\nonumber
E[ST^2] &= \frac{(\psi - E[ST_{Type1}])\cdot E[ST_{Type2}^2]}{\psi + E[ST_{Type2}] - E[ST_{Type1}]} \\
&+ \frac{ E[ST_{Type2}] \cdot E[ST_{Type1}^2] }{\psi + E[ST_{Type2}] - E[ST_{Type1}]}.
\label{Et2}
\end{align} 
The average number of packets waiting in the queue, not including the packet currently in transmission, is given by
\begin{equation}
E[N_Q] = \frac{E[ST^2]}{2{\psi}({\psi} - E[ST_{Type1}])}.
\end{equation}
Finally, the average total delay for secondary packets can be simply expressed as
\begin{equation}
E[D] = \frac{\psi E[ST_{Type2}]}{\psi + E[ST_{Type2}] - E[ST_{Type1}]} + \frac{ E[ST^2]}{2({\psi}-E[ST_{Type1}])}.
\end{equation}

\begin{figure} 
\includegraphics[width=3.4in]{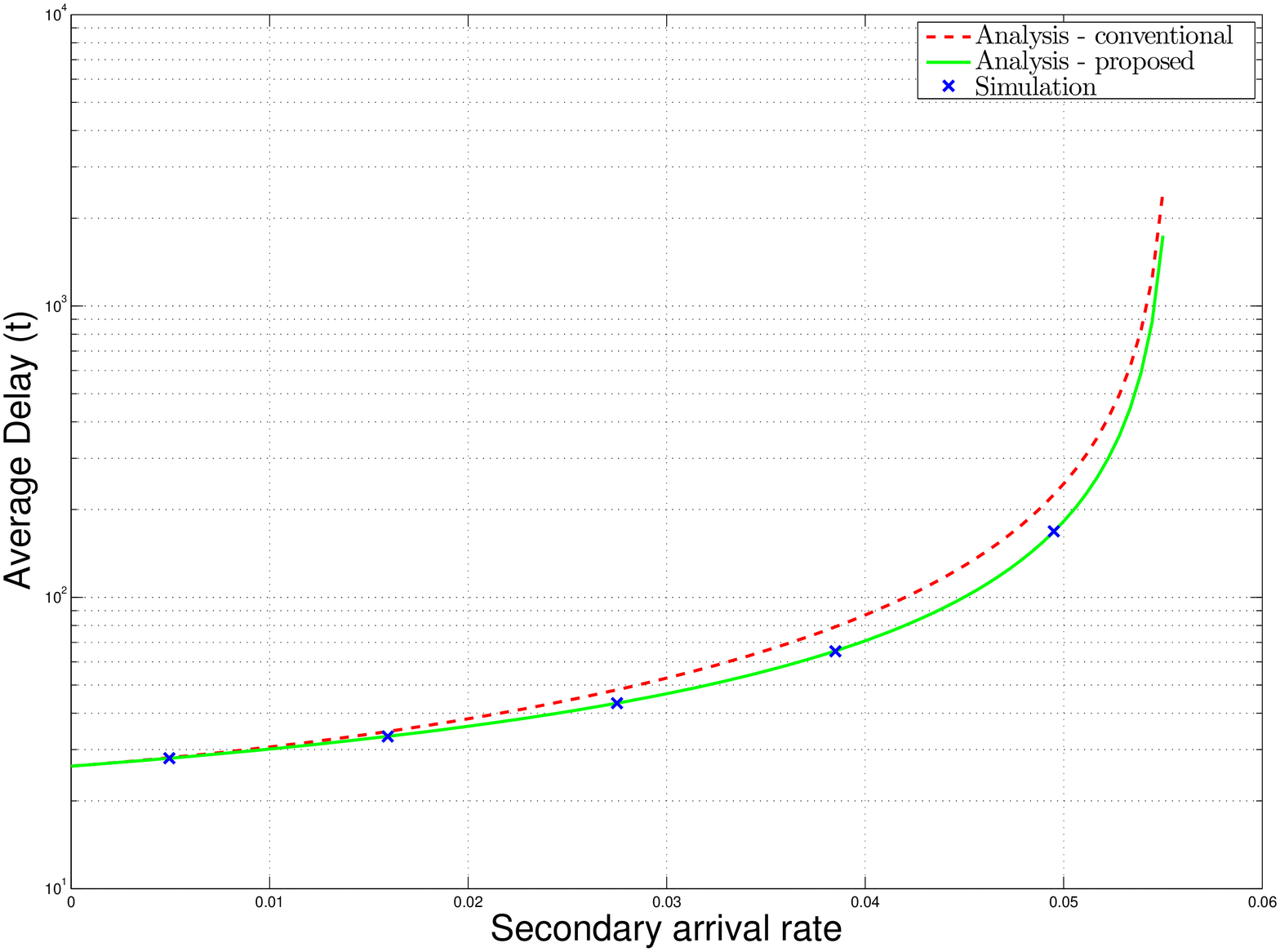}

\caption{Simulation verification for the analytical average queuing delay with continuous sensing ($T_{tr} = 3$, $\lambda = 10$, and $\mu = 2$)}
\label{ht_cont_sim}
\end{figure}


Fig. \ref{ht_cont_sim} shows the variation of average total delay for continuous sensing case against the arrival rate of data packets. The graph is based on the assumption that the average delay between packet arrival is greater than the average service time of the SU, as otherwise the queue will become unstable. The average queuing delay obtained by modeling the secondary queue with the standard M/G/1 queue is also plotted. Specifically, the conventional approach calculates the average queuing delay as $E[D] = E[ST] + \frac{ E[ST^2]}{2({\psi}-E[ST])}$, where the service time moments were simply given by Eqs. (\ref{simultaneous_eq_1}) and (\ref{Et2}). Compared with the average total delay estimated using Monte-Carlo simulation, we can clearly see that our analytical approach is more accurate. The conventional approach fails to take into account the fact that the packets in the queue are Type 1 packets.

\begin{figure} 
\includegraphics[width=3.4in]{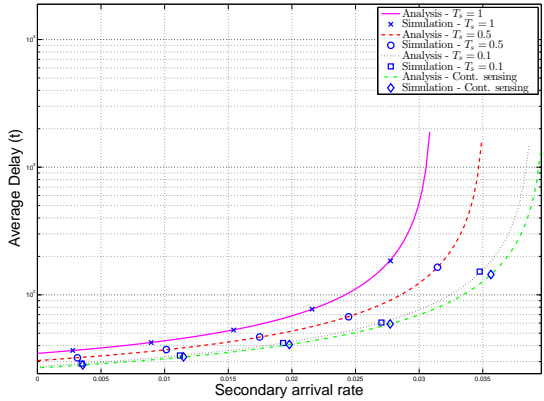}

\caption{Simulation verification for the average queuing delay ($T_{tr} = 10$, $\lambda = 3$, and $\mu = 2$).}
\label{ht_avg_delay}
\end{figure}

Fig. \ref{ht_avg_delay} shows the variation of average total delay, including the queuing delay, for periodic sensing cases with different sensing periods against the arrival rate of data packet. As the periodic sensing interval becomes small, the periodic sensing curves converge to the continuous sensing curve. The simulation results are also included to verify the analytical results obtained for the periodic sensing case.

\section{Conclusion}
\label{Conclusion}
This paper studied the extended delivery time of a data packet appearing at the secondary user in an interweave cognitive setup. Exact analytical results for the probability distribution of the EDT for a fixed-size data packet were obtained for both continuous sensing and periodic sensing. These results were then applied to analyze the expected delay of a packet at SU in a queuing setup. Simulation results were presented to verify the analytical results. These analytical results will facilitate the design and optimization of secondary systems for diverse target applications. Ongoing effort is being carried out to extend the analysis to multiple primary channels scenario with the consideration of imperfect sensing.

\section{Acknowledgment}
The authors acknowledge the contribution made by the reviewers in improving the quality of this paper.


%

\appendices 

\section{•}
\label{app_beta}
In this appendix, we derive the probability that the total SU waiting time over $k$ waiting slots is equal to $nT_s$ for the periodic sensing case, $\Pr[T_{w,k} = n T_s]$, as given in Eqs. (\ref{pr_twk}) and (\ref{beta}). $\Pr[T_{w,k} = n T_s]$ can be equivalently calculated as the probability that it takes $n$ sensing instances to find exactly $k$ times that the PU is off. Applying the result of negative binomial distribution, we can calculate $\Pr[T_{w,k} = n T_s]$ as
\begin{equation}
\Pr[T_{w,k} = n T_s] = {(1-\beta)^k}{(\beta)^{n-k}}{{n-1}\choose{k-1}},
\end{equation}
where $\beta$ is the probability that PU is on at the sensing instant. While the PU activity is modeled by a continuous-time Markov chain with transition rates given by $\frac{1}{\lambda}$ and $\frac{1}{\mu}$, $\beta$ can not be simply calculated as the stationary probability for PU on, i.e. $\frac{\lambda}{(\lambda+\mu}$. This is because with periodic sensing, the SU will sense the channel at current time instant only after the channel was sensed busy $T_s$ time period earlier. Therefore, $\beta$ should be calculated as the probability that given that PU was on at a certain point in time, the PU is again on after a time interval $T_s$, noting that there is chance that the PU turns off and then back on between two sensing instants. Based on the two-state continuous time Markov chain model for PU activity, $\beta$ can be calculated as \cite{Cinlar}
\begin{equation} 
\beta=\frac{\lambda}{\lambda+\mu}+{\frac{\mu}{\lambda+\mu}}{e^{-(\frac{1}{\lambda}+\frac{1}{\mu}){T_s}}}.
\end{equation}
The above definition is valid for any value of sensing period $T_s>0$. 
If we assume that $T_s$ is small, and the chance of the PU turning back on again before the SU senses a free channel is negligible, $\beta$ can be approximated as the probability that the exponentially distributed random variable representing the duration of a waiting slot has a value of more than $T_s$. Thus, we can approximately use $\beta={e^{\frac{-T_s}{\lambda}}}$.

\section{•}
\label{app_cont_to_per_conv}

In this appendix, we present the mathematical proof that Eq. (\ref{pr_twk}) should converge to Eq. (\ref{ftw_k}) in the limiting case, as $T_s$ approaches $0$. Note that Eq. (\ref{pr_twk}) is a PMF of negative binomial random variable, and Eq. (\ref{ftw_k}) is the PDF of an Erlang random variable. Here we reproduce the proof of convergence of negative binomial distribution to an Erlang distribution. Specifically, we will show that the MGF of $T_{w,k}$ for discrete sensing case, whose PMF is given by Eq. (\ref{pr_twk}), converges to the MGF of $T_{w,k}$, for continuous sensing case, whose PDF is given by Eq. (\ref{ftw_k}), when $T_s$ approaches $0$. The MGF of $T_{w,k}$ for the discrete case can be shown to be given by
\begin{equation}
{\cal{M}}_{T_{w,k}}(s) = \left[ \frac{(1-\beta)e^{s T_s}}{1-\beta e^{s T_s}} \right]^k.
\end{equation}
Noting that when $T_s$ is small, $\beta$ can be approximately calculated as
\begin{align}
\beta & = \frac{\lambda}{\lambda+\mu}+{\frac{\mu}{\lambda+\mu}}{e^{-(\frac{1}{\lambda}+\frac{1}{\mu}){T_s}}} \\
\nonumber
& \approx
\frac{\lambda}{\lambda+\mu} + {\frac{\mu}{\lambda+\mu}} \left( 1 - {(\frac{1}{\lambda}+\frac{1}{\mu}){T_s}} \right) = 1 - \frac{T_s}{\lambda},
\end{align}
and $\lim\limits_{T_s\to0} e^{s T_s} = 1 + s T_s$, it can be shown that
\begin{equation}
\lim\limits_{T_s\to0} {\cal{M}}_{T_{w,k}}(s) = \lim\limits_{T_s\to0} \left[ \frac{\frac{T_s}{\lambda} (1 + s T_s)}{1-(1-\frac{T_s}{\lambda}) (1+ s T_s)} \right]^k,
\end{equation}
which upon further manipulation, becomes
\begin{equation}
\lim\limits_{T_s\to0} {\cal{M}}_{T_{w,k}}(s) = \left[\frac{1}{1- \lambda s}\right]^k,
\end{equation}
which represents the MGF of the Erlang distribution with rate $\frac{1}{\lambda}$, whose corresponding PDF is defined in Eq. (\ref{ftw_k}).

\section{•}
\label{app_HT_Cont}

In this appendix, we present the calculation for the first and second moments of the EDT of packets that find PU on at the start of their service for the continuous sensing case, $E[ST_{p_{on}}]^{(c)}$ and $E[ST_{p_{on}}^2]^{(c)}$. Similar steps can be used to compute $E[ST_{p_{off}}]^{(c)}$ and $E[ST_{p_{off}}^2]^{(c)}$, as defined in Section \ref{HT_moments}.

The first moment of the EDT for the case when PU is on at the instant of packet arrival, $E[ST_{p_{on}}]^{(c)}$, is defined as
\begin{equation}
E[ST_{p_{on}}]^{(c)} = {\int}_{T_{tr}}^{\infty} t f_{{T_w},p_{on}}(t-{T_{tr}}) dt.
\end{equation}
With a change of variable $s=t-{T_{tr}}$ we get
\begin{equation}
E[ST_{p_{on}}]^{(c)} = {\int}_{0}^{\infty} (s+{T_{tr}}) f_{{T_w},p_{on}}(s) ds.
\end{equation}
After substituting Eq. (\ref{ftw_pon_summation}) and some manipulation, we arrive at
\begin{align}
\nonumber
E[ST_{p_{on}}]^{(c)} = T_{tr} & + e^{\frac{-{T_{tr}}}{\mu}} {\lambda} \sum_{k=1}^{\infty} {\frac{{T_{tr}}^{k-1} }{\mu^{k-1}{(k-1)!}{(k-1)!}}} \\
& \times {\int}_{0}^{\infty} {\frac{{s^{k}}{e^{\frac{-s}{\lambda}}}}{\lambda^{k+1}}} ds.
\end{align}
Applying the definition of the standard Gamma function, the above expression becomes
\begin{equation}
E[ST_{p_{on}}]^{(c)} = T_{tr} + e^{\frac{-{T_{tr}}}{\mu}} {\lambda} \sum_{k=1}^{\infty} {\frac{{T_{tr}}^{k-1} k}{\mu^{k-1}{(k-1)!}}}.
\end{equation}
With a change in summation variable, and after some manipulation, we get
\begin{equation}
E[ST_{p_{on}}]^{(c)} = T_{tr} + e^{\frac{-{T_{tr}}}{\mu}} {\lambda} \left[ \sum_{k=1}^{\infty} {\frac{{T_{tr}}^{k} }{\mu^{k}{(k-1)!}}} + \sum_{k=0}^{\infty} {\frac{{T_{tr}}^{k} }{\mu^{k},{k!}}} \right],
\end{equation}
which finally simplifies to
\begin{align}
\nonumber
E[ST_{p_{on}}]^{(c)} = & T_{tr} + e^{\frac{-{T_{tr}}}{\mu}} {\lambda} \left({\frac{T_{tr}}{\mu}} \right) e^{\frac{{T_{tr}}}{\mu}} +  e^{\frac{-{T_{tr}}}{\mu}} {\lambda} e^{\frac{{T_{tr}}}{\mu}}\\
& = T_{tr} + {\lambda} \left(1 + {\frac{T_{tr}}{\mu}} \right).
\label{et_on_cont}
\end{align}

The second moment, $E[ST_{p_{on}}^2]^{(c)}$ is defined as
\begin{equation}
E[ST_{p_{on}}^2]^{(c)} = {\int}_{T_{tr}}^{\infty} t^2 f_{{T_w},p_{on}}(t-{T_{tr}}) dt.
\end{equation}
With a change of variable $s=t-{T_{tr}}$, we get
\begin{align}
\nonumber
E[ST_{p_{on}}^2]^{(c)} &= {T_{tr}}^2 + 2T_{tr}(E[ST_{p_{on}}]^{(c)}-T_{tr}) \\
& + {\int}_{0}^{\infty} s^2 f_{{T_w},p_{on}}(s) ds.
\end{align}
Following the similar derivation steps for $E[ST_{p_{on}}]^{(c)}$, we get
\begin{align}
\nonumber
& E[ST_{p_{on}}^2]^{(c)} = {T_{tr}}^2 + 2T_{tr}(E[ST_{p_{on}}]^{(c)}-T_{tr}) \\
\nonumber
& + e^{\frac{-{T_{tr}}}{\mu}} {\lambda}^2 \sum_{k=0}^{\infty} \left({\frac{T_{tr}}{\mu}} \right)^{k+2} \frac{1}{k!} \\
& + 4 e^{\frac{-{T_{tr}}}{\mu}} {\lambda}^2 \sum_{k=0}^{\infty} \left({\frac{T_{tr}}{\mu}} \right)^{k+1} \frac{1}{k!} + 2 e^{\frac{-{T_{tr}}}{\mu}} {\lambda}^2 \sum_{k=0}^{\infty} \left({\frac{T_{tr}}{\mu}} \right)^{k} \frac{1}{k!}.
\end{align}
Replacing the summation by the natural exponent, we arrive at the following closed-form expression
\begin{align}
\nonumber
E[ST_{p_{on}}^2]^{(c)} = & {{\lambda}^2} \left[ {\left(\frac{T_{tr}}{\mu} \right)}^2 + 4 {\frac{T_{tr}}{\mu}} + 2 \right] \\
& + 2{\lambda}{T_{tr}}\left[ {1 + \frac{T_{tr}}{\mu}} \right] + {T_{tr}}^2.
\label{et2_on_cont}
\end{align}

\section{•}
\label{app_HT_Per}

In this appendix, we derive expressions for the first and second moments of the EDT of packets that find PU on at the start of their service for the periodic sensing case, $E[ST_{p_{on}}]^{(p)}$ and $E[ST_{p_{on}}^2]^{(p)}$. Similar steps can be used to compute $E[ST_{p_{off}}]^{(p)}$ and $E[ST_{p_{off}}^2]^{(p)}$, as defined in section \ref{HT_moments}.

When PU is on at the instant of packet arrival, the first moment of the EDT, $E[ST_{p_{on}}]^{(p)}$, is defined as
\begin{align}
E[ST_{p_{on}}]^{(p)} = E[n T_s + {T_{tr}}] = T_s E[n] + T_{tr},
\end{align}
where
\begin{align}
E[n] = & \sum_{n=1}^{\infty} n Pr[T_w,p_{on} = n T_s].
\end{align}
After substituting Eq. (\ref{pr_tw_pon_summation}) and some manipulation, we obtain
\begin{align}
\nonumber
E[n] & = {e^{\frac{-{T_{tr}}}{\mu}}} \frac{1}{(1-\beta)} \sum_{k=0}^{\infty} \left[ {{\left(\frac{{T_{tr}}}{\mu}\right)}^{k}} {\frac{k+1}{k!}} \right. \\
& \left. \times \sum_{n=k+1}^{\infty} {(1 - \beta)}^{k+2} {\beta}^{n-k-1} \frac{n!}{(n-1-k)! (k+1)!} \right].
\end{align}
Noting that the second summation is equal to $1$, the above expression can be written as
\begin{equation}
E[n] = {e^{\frac{-{T_{tr}}}{\mu}}} \frac{1}{(1-\beta)} \sum_{k=0}^{\infty} \left[ {{\left(\frac{{T_{tr}}}{\mu}\right)}^{k}} {\frac{k+1}{k!}} \right],
\end{equation}
which simplifies to
\begin{equation}
E[n] = \frac{1}{1-\beta} {\left(1 + \frac{{T_{tr}}}{\mu}\right)}.
\label{en_on}
\end{equation}
Thus, $E[ST_{p_{on}}]^{(p)}$ can be finally expressed as
\begin{equation}
E[ST_{p_{on}}]^{(p)} = {T_{tr}} + \frac{T_s}{1 - \beta} {\left(1 + \frac{{T_{tr}}}{\mu}\right)}
\end{equation}

The second moment, $E_{on}^p[t^2]$, is computed as
\begin{align}
\nonumber
E[ST_{p_{on}}^2]^{(p)} & =  E[ {(n T_s + {T_{tr}} )}^2] \\
& = {T_s}^2 E[n^2] + 2 T_s T_{tr} E[n] + {T_{tr}}^2,
\end{align}
where $E[n]$ is given in Eq. (\ref{en_on}), and
\begin{align}
\nonumber
E[n^2] & = \sum_{n=1}^{\infty} \left[ n^2 (1-\beta){\beta}^{n-1} {e^{\frac{-{T_{tr}}}{\mu}}} \right. \\
& \left. \times \sum_{k=0}^{n-1}  {\left(\frac{{T_{tr}}(1-\beta)}{\mu\beta}\right)^k}  {\frac{1}{k!}}{{n-1}\choose{k}} \right].
\end{align}
Changing the sequence of the two summations and applying $ n^2 (n-1)! = (n+1)! - n! $, we obtain
\begin{align}
\nonumber
E[n^2] = \sum_{k=0}^{\infty} & \left[ (1-\beta) {\left(\frac{{T_{tr}}(1-\beta)}{\mu\beta} \right)^{k}} {e^{\frac{-{T_{tr}}}{\mu}}}  {\frac{1}{k!}} \right. \\
\nonumber
& \left. \times \left( \sum_{n=k+1}^{\infty} {\beta}^{n-1} \frac{(n+1)!}{(n-1-k)! k!} \right. \right. \\
& \left. \left. - \sum_{n=k+1}^{\infty} {\beta}^{n-1} \frac{n!}{(n-1-k)! k!}\right) \right].
\end{align}
Using the similar manipulations for the calculation of $E_{on}^p[n]$, we obtain
\begin{align}
\nonumber
E[n^2] = \sum_{k=0}^{\infty} & \left[ {\left(\frac{{T_{tr}}}{\mu} \right)^{k}} {e^{\frac{-{T_{tr}}}{\mu}}} \right. \\
& \left. \times \left( \frac{(k+1)(k+2)}{(1-\beta)^2 {k!}} - \frac{(k+1)}{(1-\beta){k!}} \right) \right].
\end{align}
With further manipulation on the factorial terms, we arrive at
\begin{align}
\nonumber
& E[n^2] = \frac{{e^{\frac{-{T_{tr}}}{\mu}}}}{(1-\beta)^2} \left[ \sum_{k=2}^{\infty} \frac{1}{(k-2)!} {\left(\frac{T_{tr}}{\mu} \right)}^{k} \right.\\
\nonumber
& \left. + 4 \sum_{k=1}^{\infty} \frac{1}{(k-1)!} {\left(\frac{T_{tr}}{\mu} \right)}^{k} + 2 \sum_{k=0}^{\infty} \frac{1}{k!} {\left(\frac{T_{tr}}{\mu} \right)}^{k} \right]  \\
& - \frac{{e^{\frac{-{T_{tr}}}{\mu}}}}{(1-\beta)}  \left[ \sum_{k=1}^{\infty} \frac{1}{(k-1)!} {\left(\frac{T_{tr}}{\mu} \right)}^{k} + \sum_{k=0}^{\infty} \frac{1}{k!} {\left(\frac{T_{tr}}{\mu} \right)}^{k} \right],
\end{align}
which finally simplifies to
\begin{align}
\nonumber
E[n^2] = & \frac{1}{(1-\beta)^2} \left[ {\left(\frac{T_{tr}}{\mu} \right)}^2 + 4 {\frac{T_{tr}}{\mu}} + 2 \right] \\
& - \frac{1}{(1-\beta)} {\left[ \frac{T_{tr}}{\mu} + 1\right]}.
\label{en2_on}
\end{align}
Thus the second moment, $E[ST_{p_{on}}^2]^{(p)}$, can be calculated in the following closed-form expression
\begin{align}
\nonumber
E[ST_{p_{on}}^2]^{(p)} & = \frac{{T_s}^2}{{(1-{\beta})}^2} \left[ {\left(\frac{T_{tr}}{\mu} \right)}^2 + 4 {\frac{T_{tr}}{\mu}} + 2 \right] \\
& + \frac{{T_s}}{1-{\beta}} (2 T_{tr} - T_s) \cdot \left[ 1 + \frac{T_{tr}}{\mu} \right] + {T_{tr}}^2.
\label{et2_on_per}
\end{align}


\bibliographystyle{ieeetr}

\begin{thebibliography}{10}

\bibitem{Haykin}
S.~Haykin, ``Cognitive radio: brain-empowered wireless communications,'' {\em
  {IEEE} J. Sel. Areas Commun.}, vol.~23, pp.~201--220, Feb 2005.

\bibitem{Mitola}
J.~Mitola and J.~Maguire, G.Q., ``Cognitive radio: making software radios more
  personal,'' {\em {IEEE} Pers. Commun.}, vol.~6, pp.~13--18, Aug 1999.

\bibitem{Thomas}
R.~Thomas, L.~DaSilva, and A.~MacKenzie, ``Cognitive networks,'' in {\em 1st
  IEEE Int. Symp. on New Frontiers in Dynamic Spectrum Access Netw., 2005.
  DySPAN 2005. 2005}, pp.~352--360, Nov 2005.

\bibitem{Akyildiz}
I.~F. Akyildiz, W.-Y. Lee, M.~C. Vuran, and S.~Mohanty, ``Next
  generation/dynamic spectrum access/cognitive radio wireless networks: A
  survey,'' {\em Comput. Netw. J.}, vol.~50, no.~13, pp.~2127 -- 2159, 2006.

\bibitem{Islam}
M.~Islam, C.~Koh, S.~W. Oh, X.~Qing, Y.~Lai, C.~Wang, Y.-C. Liang, B.~Toh,
  F.~Chin, G.~Tan, and W.~Toh, ``Spectrum survey in {Singapore}: Occupancy
  measurements and analyses,'' in {\em Proc. 3rd Int. Conf. Cognitive Radio
  Oriented Wireless Netw. and Commun., 2008. CrownCom 2008.}, pp.~1--7, May
  2008.

\bibitem{Hamdaoui}
B.~Hamdaoui, ``Adaptive spectrum assessment for opportunistic access in
  cognitive radio networks,'' {\em {IEEE} Trans. Wireless Commun.}, vol.~8,
  pp.~922--930, Feb 2009.

\bibitem{Qianchuan}
Q.~Zhao, S.~Geirhofer, L.~Tong, and B.~Sadler, ``Opportunistic spectrum access
  via periodic channel sensing,'' {\em {IEEE} Trans. Signal Process.}, vol.~56,
  pp.~785--796, Feb 2008.

\bibitem{Qing}
Q.~Zhao, L.~Tong, A.~Swami, and Y.~Chen, ``Decentralized cognitive {MAC} for
  opportunistic spectrum access in ad hoc networks: A pomdp framework,'' {\em
  {IEEE} J. Sel. Areas Commun.}, vol.~25, pp.~589--600, April 2007.

\bibitem{Borgonovo}
F.~Borgonovo, M.~Cesana, and L.~Fratta, ``Throughput and delay bounds for
  cognitive transmissions,'' in {\em Advances in Ad Hoc Networking} (P.~Cuenca,
  C.~Guerrero, R.~Puigjaner, and B.~Serra, eds.), vol.~265 of {\em IFIP
  International Federation for Information Processing}, pp.~179--190, Springer
  US, 2008.

\bibitem{Khan}
F.~Khan, K.~Tourki, M.-S. Alouini, and K.~Qaraqe, ``Delay performance of a
  broadcast spectrum sharing network in {Nakagami-m} fading,'' {\em {IEEE}
  Trans. Veh. Technol.}, vol.~63, pp.~1350--1364, March 2014.

\bibitem{Sibomana}
L.~Sibomana, H.-J. Zepernick, H.~Tran, and C.~Kabiri, ``Packet transmission
  time for cognitive radio networks considering interference from primary
  user,'' in {\em 9th Int. Wireless Commun. and Mobile Computing Conf. (IWCMC),
  2013}, pp.~791--796, July 2013.

\bibitem{Musavian}
L.~Musavian and S.~Aissa, ``Effective capacity of delay-constrained cognitive
  radio in nakagami fading channels,'' {\em {IEEE} Trans. Wireless Commun.},
  vol.~9, pp.~1054--1062, March 2010.

\bibitem{Tran}
H.~Tran, T.~Duong, and H.-J. Zepernick, ``Delay performance of cognitive radio
  networks for point-to-point and point-to-multipoint communications,'' {\em
  EURASIP J. on Wireless Commun. and Networking}, vol.~2012, no.~1, pp.~1--15,
  2012.

\bibitem{Farraj}
A.~Farraj, S.~Miller, and K.~Qaraqe, ``Queue performance measures for cognitive
  radios in spectrum sharing systems,'' in {\em IEEE GLOBECOM Workshops (GC
  Wkshps), 2011}, pp.~997--1001, Dec 2011.

\bibitem{Jiang}
C.~Jiang, Y.~Chen, K.~Liu, and Y.~Ren, ``Renewal-theoretical dynamic spectrum
  access in cognitive radio network with unknown primary behavior,'' {\em
  {IEEE} J. Sel. Areas Commun.}, vol.~31, pp.~406--416, March 2013.

\bibitem{Gaaloul}
F.~Gaaloul, H.-C. Yang, R.~Radaydeh, and M.-S. Alouini, ``Switch based
  opportunistic spectrum access for general primary user traffic model,'' {\em
  {IEEE} Wireless Commun. Lett.}, vol.~1, pp.~424--427, October 2012.

\bibitem{Liang}
Z.~Liang and D.~Zhao, ``Quality of service performance of a cognitive radio
  sensor network,'' in {\em Proc. IEEE Int. Conf. Commun. (ICC), 2010},
  pp.~1--5, May 2010.

\bibitem{J_Wang}
X.~Li, J.~Wang, H.~Li, and S.~Li, ``Delay analysis and optimal access strategy
  in multichannel dynamic spectrum access system,'' in {\em Proc. Int. Conf.
  Computing, Netw. and Commun. (ICNC), 2012}, pp.~376--380, Jan 2012.

\bibitem{Kandeepan}
S.~Kandeepan, C.~Saradhi, M.~Filo, and R.~Piesiewicz, ``Delay analysis of
  cooperative communication with opportunistic relay access,'' in {\em Proc.
  IEEE 73rd Veh. Technol. Conf. (VTC Spring), 2011}, pp.~1--5, May 2011.

\bibitem{Li_Han}
H.~Li and Z.~Han, ``Queuing analysis of dynamic spectrum access subject to
  interruptions from primary users,'' in {\em 5th Int. Conf. Cognitive Radio
  Oriented Wireless Netw. Commun. (CROWNCOM), 2010 Proc.}, pp.~1--5, June 2010.

\bibitem{Kahvand}
M.~Kahvand, M.~Soleimani, and M.~Dabiranzohouri, ``Channel selection in
  cognitive radio networks: A new dynamic approach,'' in {\em Communications
  (MICC), 2013 IEEE Malaysia International Conference on}, pp.~407--411, Nov
  2013.

\bibitem{Namanya}
A.~Namanya and J.~Pagna-Disso, ``Performance modelling and analysis of the
  delay aware routing metric in cognitive radio ad hoc networks,'' in {\em
  Wireless and Mobile Networking Conference (WMNC), 2013 6th Joint IFIP},
  pp.~1--8, April 2013.

\bibitem{W_Wang}
C.-W. Wang and L.-C. Wang, ``Analysis of reactive spectrum handoff in cognitive
  radio networks,'' {\em {IEEE} J. Sel. Areas Commun.}, vol.~30,
  pp.~2016--2028, November 2012.

\bibitem{W_Wang2}
L.-C. Wang, C.-W. Wang, and F.~Adachi, ``Load-balancing spectrum decision for
  cognitive radio networks,'' {\em {IEEE} J. Sel. Areas Commun.}, vol.~29,
  pp.~757--769, April 2011.

\bibitem{MacKay}
D.~J.~C. MacKay, ``Fountain codes,'' {\em IEE Proc.-Commun.}, vol.~152,
  pp.~1062--1068, Dec 2005.

\bibitem{Castura}
J.~Castura and Y.~Mao, ``Rateless coding over fading channels,'' {\em IEEE
  Commun. Lett.}, vol.~10, pp.~46--48, Jan 2006.

\bibitem{Adan}
I.~Adan and J.~Resing, {\em Queueing Theory}.
\newblock Eindhoven University of Technology. Department of Mathematics and
  Computing Science, 2001.

\bibitem{Liu}
Q.~Liu, X.~Wang, and Y.~Cui, ``Robust and adaptive scheduling of sequential
  periodic sensing for cognitive radios,'' {\em {IEEE} J. Sel. Areas Commun.},
  vol.~32, pp.~503--515, March 2014.

\bibitem{Mariani}
A.~Mariani, K.~Sithamparanathan, and A.~Giorgetti, ``Periodic spectrum sensing
  with non-continuous primary user transmissions,'' {\em {IEEE} Trans. Wireless
  Commun.}, vol.~PP, no.~99, pp.~1--1, 2014.

\bibitem{B_Wang}
B.~Wang, Z.~Ji, and K.~Liu, ``Primary-prioritized markov approach for dynamic
  spectrum access,'' in {\em 2nd IEEE Int. Symp. on New Frontiers in Dynamic
  Spectrum Access Netw., 2007. DySPAN 2007}, pp.~507--515, April 2007.

\bibitem{Zhang}
Y.~Zhang, ``Spectrum handoff in cognitive radio networks: Opportunistic and
  negotiated situations,'' in {\em IEEE Int. Conf. on Comm., 2009. ICC '09.},
  pp.~1--6, June 2009.

\bibitem{Akyildiz2}
I.~Akyildiz, W.-Y. Lee, M.~C. Vuran, and S.~Mohanty, ``A survey on spectrum
  management in cognitive radio networks,'' {\em IEEE Communications Magazine},
  vol.~46, pp.~40--48, April 2008.

\bibitem{Cabric}
D.~Cabric, S.~Mishra, and R.~Brodersen, ``Implementation issues in spectrum
  sensing for cognitive radios,'' in {\em Conference Record of the
  Thirty-Eighth Asilomar Conference on Signals, Systems and Computers, 2004.},
  vol.~1, pp.~772--776 Vol.1, Nov 2004.

\bibitem{Molisch}
A.~Molisch, N.~Mehta, J.~S. Yedidia, and J.~Zhang, ``Performance of fountain
  codes in collaborative relay networks,'' {\em {IEEE} Trans. Wireless
  Commun.}, vol.~6, pp.~4108--4119, November 2007.

\bibitem{Cinlar}
E.~Cinlar, {\em Introduction to stochastic processes}.
\newblock Englewood Cliffs, NJ: Prentice-Hall, 1975.

\bibitem{Boucherie}
R.~Boucherie and N.~van Dijk, {\em Queueing Networks: A Fundamental Approach}.
\newblock International Series in Operations Research \& Management Science,
  Springer, 2010.

\end{thebibliography}

\end{document}